\documentstyle[aps,prl,multicol,epsf]{revtex}

\begin{document}
\title{Enumeration of States in a Periodic Glass}
\author{P. Chandra$^1$, L.B. Ioffe$^{2,3}$ and D. Sherrington$^{4}$}
\address{$^1$NEC Research Institute, 4 Independence Way, Princeton NJ
08540}
\address{$^2$Landau Institute for Theoretical Physics, Moscow, RUSSIA}
\address{$^3$Department of Physics, Rutgers University, Piscataway, NJ
08855}
\address{$^4$Department of Physics, Theoretical Physics, Oxford University, \\1
Keble Road, Oxford, OX1 3NP, United Kingdom}
\maketitle

\begin{abstract}
We present an analytic enumeration of the metastable states, $N_s$, in a
periodic long-range Josephson array frustrated by a transverse field. We
find that the configurational entropy, $S_{conf} \equiv \ln N_s$, is
extensive and scales with frustration, confirming that the non-random system
is glassy. We also find that $S_{conf}$ is different from that of its
disordered analogue, despite that fact that the two models share the same
dynamical equations.
\end{abstract}

%
%

\begin{multicols}{2}
The problem of vitrification has been studied for many years
and remains a topic of much current interest \cite{Angell95}. Recently
there has been significant progress in our 
{\em conceptual} understanding of glass formation in the absence of
quenched disorder due to studies of exactly soluble models
\cite{Bouchaud94,Marinari94,Cugliandolo95,Franz95,Chandra95,Bouchaud97}. Many
features of these regular microscopic systems are the same as those of known
disordered Hamiltonians \cite{Kirkpatrick87}. More specifically, the
dynamical equations of these {\sl periodic} glasses are identical to those
of (intrinsically random) spherical spin glass models; many signatory
properties such as history-dependence and ageing follow from these equations.
These results suggest that one can study periodic glasses via a mapping to
disordered ones for which analytical tools are well-developed \cite
{Bouchaud96,Parisi97}. Here, however, we show that some physical properties
of periodic and disordered models which share the same dynamical equations
are different. In particular we have calculated the number of metastable
states, $N_s$, with different ordering of the superconducting phases in a periodic long-range Josephson array. We find that the
configurational entropy, $S_{conf} \equiv \ln N_s$, is extensive proving
that this system is indeed in a glassy phase. However, $S_{conf}$ in this
model is {\sl distinct} from the configurational entropy of its disordered
counterpart, despite the fact that they share the same dynamical equations.

We shall consider the following periodic model, which has the advantage that
it can be studied {\sl both} analytically \cite{Chandra95,Chandra96} and
experimentally \cite{Chandra97a,Tinkham97}. The proposed array is a stack of
two mutually perpendicular sets of $N$ parallel wires with Josephson
junctions at each node that is placed in an external tranverse field, $H$.
The classical thermodynamic variables of this system are the superconducting
phases associated with each wire. Here we shall assume that the Josephson
couplings are sufficiently small so that the induced fields are negligible
in comparison with $H$. We can therefore describe the array by the
Hamiltonian 
\begin{equation}
{\cal H} = - \sum_{m,n}^{2N} z_m^{*}\ {\cal J}_{mn} \ z_n  \label{H}
\end{equation}
where ${\cal J}_{mn}$ is the coupling matrix 
\begin{equation}
\hat{{\cal J}} = \left( 
\begin{array}{cc}
0 & \hat{J} \\ 
\hat{J}^\dagger & 0
\end{array}
\right)  \label{J}
\end{equation}
with $J_{jk} = \frac{J_0}{\sqrt{N}} \exp(2\pi i \alpha jk /N)$ and $1 \!
\leq \! (j,k) \! \leq \! N$ where $j(k)$ is the index of the horizontal
(vertical) wires; $z_m = e^{i\phi_m}$ where $1 \leq m \leq 2N$ and the $%
\phi_m$ are the phases associated with the superconducting order parameters
of the $2N$ wires. Here we have introduced the flux per unit strip, $\alpha
= NHl^2/\Phi_0$, where $l$ is the inter-node spacing and $\Phi_0$ is the
flux quantum; the normalization has been chosen so that $T_G$ does not scale
with $N$.

Because every horizontal (vertical) wire is linked to every vertical
(horizontal) wire, the number of nearest neighbors in this model is $N$; we
can therefore study it with a mean-field approach. Such an analysis of the
thermodynamic properties indicates that at sufficiently low temperatures the
paramagnetic phase becomes unstable with $\alpha N$ modes becoming
simultaneously degenerate\cite{Chandra95}. We speculated previously that any
linear combination of these modes would result in a metastable state at
lower temperatures leading to $S_{conf} \sim \alpha N$. The results of the
explicit calculation described below confirm this conjecture.

In the limit of small transverse field $(\frac{1}{N} < \alpha < 1)$, the
dynamical equations of the periodic array are identical to those of the $p=4$
(disordered) spherical spin glass model \cite{Chandra96}. These equations
indicate that all metastable states are formed at the transition, with no
further subdivision occurring at lower temperatures \cite{Chandra97b}. This
conclusion agrees with that found from a direct study of the TAP solutions
of the $p=4$ spherical spin glass model \cite{Kurchan93}. It is thus
sufficient to calculate the number of states, $N_s$, in the periodic model
at zero temperature. Because there is no average over disorder in 
this regular array, there is no distinction between $\ln \langle N_s
\rangle$
and $\langle \ln N_s \rangle$ as occurs for intrinsically random
systems,
and hence no corresponding problems of relevance or interpretation of
$\ln \langle N_s \rangle$ that one usually calculates instead of the
more physical  $\langle \ln N_s \rangle$.

At zero temperature the defining characteristic of a metastable state is
that each spin should be parallel to its associated molecular field \cite
{Tanaka80a}. This physical condition results in a highly nonlinear equation
which determines the number of states. Previous studies of disordered $xy$
spin systems indicate that the crucial nonlinearities appear in the
expression for the amplitude, due to the quasi-linear nature of the phase
component. More specifically, the configurational entropy for the $xy$
Sherrington-Kirkpatrick model\cite{Tanaka80b} is different from that of its
Ising counterpart by a numerical factor of $O(1)$. This observation allows
one to reduce the array problem to that of Ising spins, $s_m = \pm 1$,
thereby simplifying the technical presentation. In this case, the number of
states is given by 
\begin{eqnarray}
N_s & \equiv & {\rm {Tr}}_s \prod_m \theta\left (s_m \sum_n {\cal J}_{mn} s_n
\right)  
\label{NsDefinition} \\
& = & \int^{+\infty}_{-\infty} {\cal D} \phi \hspace{0.05in} {\rm {Tr}}_s 
\hspace{0.05in} e^{i \sum_{mn} {\cal J}_{mn} {\cal A}_{mn} s_m s_n}
\label{1}
\end{eqnarray}
where we have used an integral representation of the $\theta$-function, $%
{\cal D}\phi \equiv {\prod_m \frac{{\it {d}\phi_m}}{2\pi i\phi_m}}$ and $%
{\cal A}_{mn} \equiv (\phi_m + \phi_n)$.

We use the cumulant expansion to determine the integral (\ref{1}).
Since only terms with an even power of each spin variable
contributes to this sum, it can be represented as a closed loop
product of $(J{\cal A})_{ij}$ matrices.  Moreover the structure
of the periodic array is such that only loops containing
an even product of these matrices are possible \cite{Chandra95,Vinokur87}.
For the explicit calculation we shall 
need the moments of the couplings $\langle J^{2p} \rangle$.
We know
from previous work on the array \cite{Chandra95,Vinokur87} that 
\begin{eqnarray}
\langle J^{2p} \rangle & = & \left(\frac{1}{N^{2p-1}}\right) \lim_{N
\rightarrow \infty} \sum_{i_1,j_1,i_2...j_p} J_{i_1 j_1}J_{j_1 i_2}\hspace{%
0.01in}...\hspace{0.01in}J_{j_p i_1}  \nonumber \\
& = & \frac{1}{\alpha^{p-1}N^{2p-1}}  \label{2}
\end{eqnarray}
where $J_0 \equiv 1$; this result can be understood physically since only
horizontal (vertical) wires separated by $d < l/\alpha$ contribute
coherently to this sum independent of disorder. We emphasize that so far our
calculation can be applied to {\sl both} periodic and random arrays; the
only difference is that in the latter case the couplings are averaged over
disorder.

\begin{figure}
\begin{minipage}[c]{80mm}
\centerline{\epsfxsize=6.5cm \epsfbox{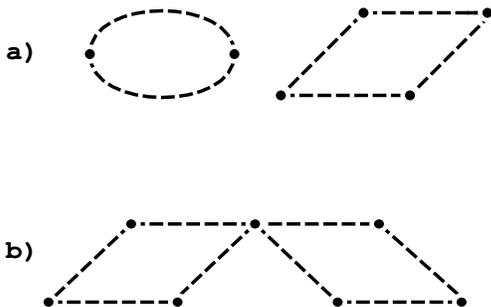}} 
\caption{Diagrams for the spin trace: (a) Closed non-intersecting
loops that give the main contribution to $N_s$ in the limit $\alpha \ll 1$
and (b) Intersecting loops that are next order in $\alpha$.} 
\end{minipage}
\end{figure}

The summation over spin variables in (\ref{1}) can be graphically
represented by the sum of diagrams with closed loops, shown in Fig. 1.
Diagrams with intersecting loops (see Fig. 1) are small in $\alpha$ due to
the structure of the moments (\ref{2}); thus we can neglect such diagrams
for $\alpha < 1$ which simplifies the problem. Furthermore each closed loop
diagram, (see Fig. 1a), makes a contribution $K^{2p} = \int {\cal D}\phi \ 
{\rm {Tr}\ ({\cal J A})^{2p}}$ to (\ref{1}) where $p$ refers to the number
of nodes in the graph. The integration over the phase variables makes all
sites equivalent, so that $K^{2p} = \langle {\cal J}^{2p}\rangle \int {\cal D%
} \phi \ {\rm {Tr}\ {\cal A}^{2p}}$. Summing all the loop diagrams and
rescaling $\phi \rightarrow \sqrt{\alpha} \phi$, we have 
\begin{equation}
{N_s} = 2^N \int {\cal D}\phi \hspace{0.02in} e^{\alpha N {\cal S} (\phi)} 
\hspace{0.12in}, \hspace{0.12in} {\cal S}(\phi) = \sum_p \frac{(-1)^{2p}}{%
(2p)N^{2p}} {\rm {Tr} \hspace{0.02in}{\bf A}^{2p}.}  \label{4}
\end{equation}
More explicitly, because ${\cal A}_{ij} = (\phi_i + \phi_j)$ 
\begin{equation}
\mbox  
{Tr}\ {\bf A}^{n} \equiv \sum_{i_1,\ldots i_n} (\phi_{i_1}+\phi_{i_2})
(\phi_{i_2}+\phi_{i_3}) \ldots (\phi_{i_n}+\phi_{i_1}).  \label{TrA^n}
\end{equation}
In each term of the resulting polynomial each index is represented zero, one
or at most two times; the resulting contributions to $\frac{1}{N^n} {\rm Tr}%
{\bf A}^n$ are $\frac{1}{N} \sum_i 1 \equiv 1$, $\frac{1}{N} \sum_i \phi_i $
and $\frac{1}{N} \sum_i \phi_i^2$ respectively. Therefore $\frac{1}{N^{2p}} 
{\rm Tr}{\bf A}^{2p}$ is a function of only $t \equiv \frac{1}{N} \sum \phi$
and $u \equiv \frac{1}{N} \sum \phi^2$. In order to calculate $\mbox{Tr} 
{\bf A}^n$ explicitly, we separate the two terms in which each index appears
once and only once from those in which some indices are repeated and others
do not appear at all. Furthermore, it is convenient to compute the second term
via its derivative: 
\begin{equation}
\mbox{Tr}\ {\bf A}^n = 2 t^n + \int_0^{u} \frac{\partial \mbox{Tr} {\bf A}^n%
} {\partial u} d u.  \label{TrA^n_1}
\end{equation}
Explicitly the derivative in the second term is given by 
\begin{equation}
\frac{\partial \mbox{Tr}\ {\bf A}^n} {\partial u} = n N C_{n-2}
\end{equation}
\begin{equation}
C_{n} = \sum_{i_1,\ldots i_{n+1}} (\phi_{i_1}+\phi_{i_2})
(\phi_{i_2}+\phi_{i_3}) \ldots (\phi_{i_n}+\phi_{i_{n+1}}).  \label{C_n}
\end{equation}
Here $C_n$ can be determined recursively, once again separating into terms
which involve only distinct and repeated indices. Explicit computation of
the sum in equation (\ref{4}), involving some algebra, yields 
\begin{equation}
{\cal S} = - \frac{1}{2} \ln \left\{(1 + t^2)^2 + 2u(1 - t^2) + u^2 \right\}.
\label{5}
\end{equation}
Because ${\cal {S}}$ depends on $\phi$ solely via $t$ and $u$ defined above,
the $N$-dimensional integral over $\phi$ in (\ref{4}) can be conveniently
calculated by including two additional integrals over $t$ and $u$, 
and the factors $\delta\left(t-\frac{1}{N}\sum \phi\right)$ and $\delta\left(u - \frac{1}{N}%
\sum \phi^2\right)$.

\begin{figure}
\begin{minipage}[c]{80mm}
\centerline{\epsfxsize=7.5cm \epsfbox{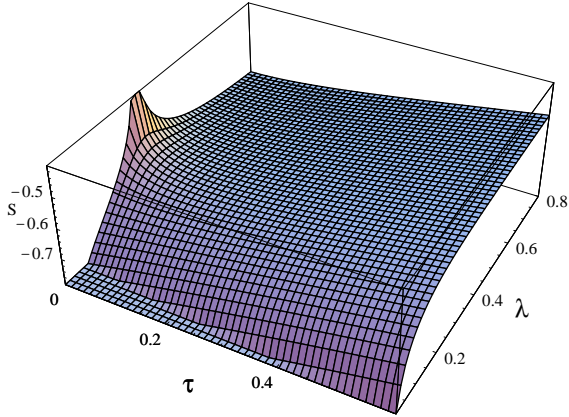}}
\caption{${\cal S}(\tau,\lambda)$ for a typical $\alpha$
($\alpha =0.1$) which displays the saddle-point.}
\end{minipage} 
\end{figure}

We may use the integral representation of the $\delta$-functions and
integrate over each $\phi_i$ independently; the result can be concisely
presented using the function $\Theta(x) \equiv \frac{1}{2} \left\{ 1 + {\rm {%
Erf}(\frac{x}{2})}\right\}$ so that the integral in (\ref{4}) becomes 
\begin{equation}
{N_s} = 2^N \int^{i\infty}_{-i\infty} \! d \lambda \!
\int^{\infty}_{-\infty} \! dt \
du\ d\tau \hspace{0.05in} e^{N\left\{\alpha {\cal {S}} + i\tau t + \lambda u
+ \ln \Theta\left(\frac{\tau}{\sqrt{\lambda}}\right)\right\}}  
\label{6}
\end{equation}
where ${\cal {S}}$ is given by (\ref{5}). The factor of $N$ in the
exponential of the integral in (\ref{6}) allows us to evaluate this integral
by a saddle-point approximation. We shift our contour in $t$ by $i\tilde{t}$
to look for a solution, since on physical grounds we expect the saddle-point
to be real. This is indeed the case, as displayed in Figure 2, and the
result is 
\begin{equation}
S_{conf} \equiv \ln N_s = \frac{N\alpha}{2} \ln \frac{1}{2\alpha}.  \label{7}
\end{equation}
for $\alpha \ll 1$. This equation for the configurational entropy is the
main result of this paper. It indicates that $S_{conf}$ is extensive, as
expected for a glass.  Furthermore  $S_{conf}$ is proportional to $\alpha N$,
consistent with the fact that there is vanishing frustration in the limit
of $\alpha \le \frac{1}{N}$. Physically, we note that the
configurational
entropy in (\ref{7}) is proportional to the effective number of
spins $\alpha N$ in the periodic model; in this array,
phases (and hence spins) are correlated
on a length-scale 
$\frac{l}{\alpha}$ where $l$ is the internode spacing.
Eq. (\ref{7}) also concurs with a conjecture
based on a previous stability analysis for the high temperature paramagnetic
phase \cite{Chandra95}.

The calculation above can be generalized to find the number of states in
which all molecular fields are greater than a threshold value, $h_{th}$. For
small $h_{th}$ the associated configurational entropy decreases as $\delta
S_{conf} \sim - N \alpha^{1/2} h_{th}$. We note that {\sl both} $S_{conf}$
and $\delta S_{conf}$ are proportional to $N$. Empirically we believe that
states in which some spins experience a very small molecular field are only
marginally stable. Thus this perturbative change in the configurational
entropy induced by finite $h_{th} \ll 1$ implies that in the thermodynamic
limit the number of marginal states is {\sl extensive}.

We can also extend this calculation to determine the number of states
as a function of their energy. To do this we introduce an additional
$\delta(N E-\sum_{mn} {\cal J}_{mn} s_m s_n)$ function in the definition
(\ref{NsDefinition}) which extracts only states with energy $E$. Using an
integral representation of this $\delta$ function with an additional
variable $\mu$ we see that ${\cal D} \phi$ in Eq (\ref{1}) becomes 
${\cal D}\phi = \Pi_m \frac{d \phi_m}{2\pi i (\phi_m+\mu)}$ and the
exponential acquires an additional contribution $i N \mu E$. Finally in the
final integral (\ref{6}) the exponential acquires the term $i N E
\mu$ and function $\Theta(\tau/\sqrt{\lambda})$ is replaced by 
\begin{eqnarray}
\tilde{\Theta}(\tau,\lambda) &=& \int \frac{d \phi}{2\pi i (\phi+\mu)}
e^{-\lambda \phi^2 + i \tau \phi} 
\nonumber \\
&=& e^{-\mu^2 \lambda - i \tau \mu} 
\Theta \left( \frac{\tau - 2i \mu \lambda}{\sqrt{\lambda}} \right)
\label{NewTheta}
\end{eqnarray}
We evaluate the resulting integral by a saddle-point approximation.
Three out of the five resulting saddle-point equations 
can be solved analytically;
the remaining two contain an error function and must be solved
numerically.     
\begin{figure}
\begin{minipage}[c]{80mm}
\centerline{\epsfxsize=7.5cm \epsfbox{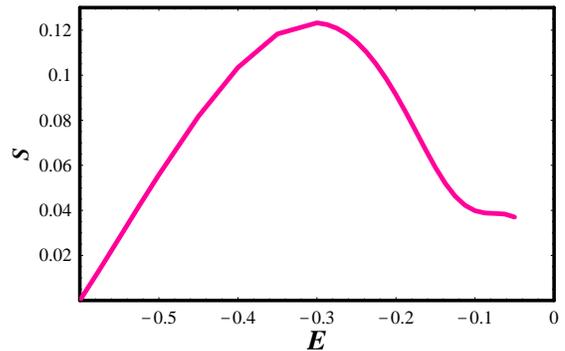}}
\caption{${\cal S}_{conf}(E)$ for $\alpha=0.1$}
\end{minipage} 
\end{figure}

Our numerical solution  for the configurational entropy as a function
of energy is displayed in Figure 3 for $\alpha = 0.1$.
Qualitatively it is similar to the behavior of $S_{conf}(E)$ for
spin glasses with intrinsic 
randomness\cite{Tanaka80a,Tanaka80b,Crisanti95}. However its
analytical structure is {\sl distinct}, specifically from that of the $p=4$
(disordered) spherical model \cite{Crisanti95}.
Furthermore all states displayed in Fig. 3 are stable,
whereas in the $p$-spin model all states above
a certain threshhold energy are unstable.
We note, however, that here
we are only considering stability with respect to single-spin flips;
by contrast in the $p$-spin model stability is defined with
respect to continuous deformations.

The result (\ref{7}) was derived in the limit of small $\alpha$, and we have
also checked the qualitative validity of our conclusion for $\alpha = 1$
with $xy$ spins using a different method than that discussed above. We
emphasize that the $\alpha = 1$ case is very special, since in this instance 
$J_{ij}$ is a unitary matrix. Furthermore we note that for $\alpha = 1$ the
coupling matrix $J_{ij}$ is identical to a discrete Fourier transform; then
the condition that the Fourier transform is flat defines a stable spin
configuration. It is convenient to write this condition as $B_r \equiv {%
\sum_{k=0}^{N} z^*_k z_{k+r}}$ equals zero for $r \neq 0$. Then we obtain the
equation for the number of states 
\begin{equation}
{N_s} = \int \prod_{r=1}^N dz_r\hspace{0.05in} dz^*_r \hspace{0.05in} \frac{%
\delta(|z_r| -1)}{2\pi} \hspace{0.05in} \delta(B_r) \hspace{0.05in} \det 
\frac{\partial B}{\partial z}.  \label{8}
\end{equation}
Since the $B_r$ are pseudo-random variables with $\langle B^2\rangle = N$ we
may assume that they obey a Gaussian distribution with $P(B) \approx \frac{1%
}{2\pi N} e^-\frac{|B^2|}{N}$. The determinant in (\ref{8}) can be evaluated
exactly by noting that $\frac{\partial B_k}{\partial z} = z^*_{i+k}$
implying that for $z$'s that satisfy the condition $B_r = 0$ all rows are
orthogonal. The length of each row is $N$ so that the determinant $\vert 
\frac{\partial B}{\partial z}\vert = N^N$. Therefore the average over $B$ in
(\ref{8}) results in 
\begin{equation}
S_{conf} \equiv \ln N_s = N \ln \sqrt{\frac{N}{2\pi}}.  \label{9}
\end{equation}
This expression was derived in the limit of $N \gg 1$; from its form it is
clear that this asymptotic result can be attained only at numerically large $%
N$ (at least $N \gg 2 \pi$). In order to check the behavior of $S_{conf}$
for $\alpha = 1$ at moderate $N$, we have performed a direct numerical
minimization of the Hamiltonian in (\ref{H}); the results, displayed in Fig.
3, clearly indicate that the configurational entropy is extensive. We expect
that $S_{conf}$ crosses over from $S_{conf} \approx N \ln2$ (obtained
numerically as shown in Fig. 3) to the analytical result (\ref{9}) at $N
\approx 20$; however from a practical standpoint this crossover is
inaccessible numerically due to the extensive nature of the ground-state
manifold.

To summarize our results, we have calculated the number of states in a
periodic glass and have found that the configurational entropy, $S_{conf} =
\ln N_s$, is proportional to $\alpha N$ (Eq. (\ref{7})). It is thus
extensive but is {\sl different} from that of the $p=4$ (disordered)
spherical spin glass model \cite{Bouchaud97} despite the fact that their
dynamical equations are {\sl identical} \cite{Chandra96,Chandra97b}.
Furthermore, the complexities (the number of states as a function of
energy) of these two models have different functional forms. It
appears that rather different microscopic models can have the same dynamical
behavior, as evidenced by the fact that the parameter $\alpha$ (for $\alpha
< 1$) does not affect the rescaled response. As an aside, we note that $%
S_{conf}$  for the disordered array is
also distinct from that of the Sherrington-Kirkpatrick model, even though
they share the same dynamics \cite{Vinokur87}; however $S_{conf}$ for the
periodic and the random networks are identical, which may be coincidental.
These qualitative results remain valid for $\alpha = 1$, thus suggesting
that they are independent of our approximation given the absence of
commensurability. We note, however, that the dynamical equations change
their form at $\alpha \sim 1$, and become similar to those of disordered
spherical spin glass models with interactions that are sums of couplings
with $p \ge 4$; qualitatively this does not seem to affect the number of
states.

\begin{figure}
\begin{minipage}[c]{80mm}
\centerline{\epsfxsize=8.5cm \epsfbox{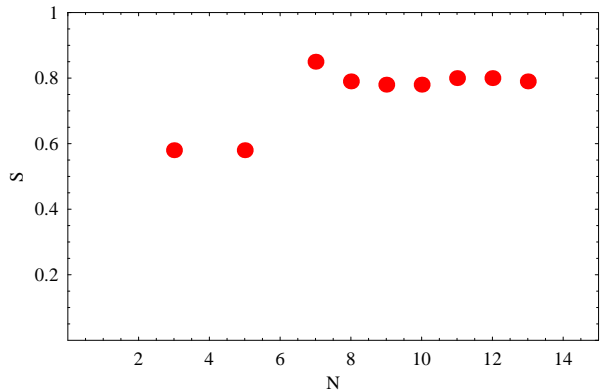}}
\caption{Numerical results for the configurational entropy, $s\equiv
S_{conf}/N$, for $\alpha = 1$ as a function of $N$. Note that for $N <
12$ the possible error is less than the size of the points.}
\end{minipage} 
\end{figure}

The fact that the periodic long-range array and the $p=4$ 
spherical (disordered)
spin glass have different configurational entropies
suggests that the detailed structure of
their respective phase spaces, 
e.g. barriers, basins of attraction, is quite
distinct.  
This is not inconsistent with previous work which indicates
that these two models share the same 
dynamical equations\cite{Chandra96,Chandra97b}.
The latter were derived in the thermodynamic limit ($N
\rightarrow \infty$),
and do not describe transitions between metastable
states that might occur at finite $N$.
Thus these equations only probe the system's
responses in the vicinity of a typical state where it is trapped,
and are insensitive to the total number of
metastable configurations.
The distinction between the configurational entropies
of the periodic and disordered models indicates differences in their
physical
properties at finite $N$ and, most likely, in the characteristics of analogous
finite-range systems. We therefore believe that a detailed study of
the 
structure of their respective phase spaces 
could give insight about the physics of finite-range problems and
about differences between disordered and structural glasses.

Experimentally states are accessed with some weighing factor which is
ignored by the quantity $N_s$ calculated here; for example in thermal
equilibrium they would be weighted with a Boltzmann factor, whereas in
a rapid quench they would be weighted according to the size of their
basins of attraction. Furthermore it seems likely \cite{Chandra98}
that the distribution of the basins of attraction is broad; if so,
some states would have  weights much larger than others in a rapid
quench. Nonetheless 
we believe that the main conclusion of this paper remains valid;
the
number of states can be very different for systems which share the same dynamical equations.

We thank D.S. Fisher for raising the issue of stability of the states. P.C.
acknowledges the hospitality of the Department of Physics and
Brasenose College during a sabbatical leave at Oxford. L.I. is grateful to
the same Physics Department and to New College (Oxford) for their support
during the period of this research. The work performed at Oxford University
was partially funded by the UK EPSRC under grant DR 8729.

\end{multicols}


\end{document}